\documentclass[aps,prd,tightenlines,nofootinbib,twocolumn,10pt]{revtex4-1}

\usepackage[english]{babel}

\usepackage[utf8x]{inputenc}
\usepackage[T1]{fontenc}
\usepackage{ucs}
\usepackage{microtype}
\usepackage{newtxtext,newtxmath}
\usepackage{color}

\usepackage{url}
\usepackage{hyperref}

\usepackage{xspace}
\usepackage{lastpage}
\usepackage{stmaryrd}

\usepackage{graphicx}
\usepackage{empheq}
\usepackage{ifthen}
\usepackage{tensor}
\usepackage{paralist}

\usepackage{amsmath}
\usepackage{amsfonts}
\usepackage{amssymb}
\usepackage{mathrsfs}
\usepackage{extarrows}
\usepackage{verbatim}
\usepackage{upgreek}
\usepackage{wasysym}

\usepackage{natbib}

\newcommand\define{\equiv}

\newcommand\vect[1]{\boldsymbol{#1}}

\newcommand\ex[1]{\mathrm{e}^{#1}}

\newcommand\e[1]{_{\text{#1}}}
\newcommand\h[1]{^{\text{#1}}}
\newcommand\U[1]{\:\mathrm{#1}}

\newcommand{\dd}{\mathrm{d}}

\renewcommand\lim[2]{\underset{ #1 \rightarrow #2 }{ \mathrm{lim} } \,}

\newcommand{\delimiters}[4][]{
\ifthenelse{ \equal{#1}{1} }{  #2 #3 #4  }
					{ \ifthenelse{\equal{#1}{2}}{ \big#2 #3 \big#4 }
						{ \ifthenelse{\equal{#1}{3}}{ \Big#2 #3 \Big#4 }
							{ \ifthenelse{\equal{#1}{4}}{ \bigg#2 #3 \bigg#4 }
								{ \ifthenelse{\equal{#1}{5}}{ \Bigg#2 #3 \Bigg#4 }
									{ \left#2 #3 \right#4 }
								}
							}
						}
					}
													}

\newcommand{\pa}[2][]{\delimiters[#1]{(}{#2}{)}}
\newcommand{\pac}[2][]{\delimiters[#1]{[}{#2}{]}}

\newcommand{\abs}[2][]{\delimiters[#1]{|}{#2}{|}}

\newcommand{\cell}{\ensuremath{\mathcal{C}}}
\newcommand{\disc}{(D)}
\newcommand{\cont}{(C)}
\newcommand{\oone}{\raisebox{-1.5pt}{\textcircled{\raisebox{1.5pt} {\tiny 1}}}}
\newcommand{\otwo}{\raisebox{-1.5pt}{\textcircled{\raisebox{1.5pt} {\tiny 2}}}}

\newcommand{\aap}{A{\&}A}
\newcommand{\mnras}{MNRAS}
\newcommand{\jcap}{JCAP}
\newcommand{\physrep}{Phys. Rep.}

\begin{document}

\title{Cosmic backreaction and Gauss's law}

\author{Pierre Fleury}
\email{pierre.fleury@uct.ac.za}
 
\affiliation{Department of Mathematics and Applied Mathematics, University of Cape Town,
Rondebosch 7701, Cape Town, South Africa\\
Department of Physics and Astronomy, University of the Western Cape,
Robert Sobukwe Road, Bellville 7535, South Africa
}

\begin{abstract}
Cosmic backreaction refers to the general question of whether a homogeneous and isotropic cosmological model is able to predict the correct expansion dynamics of our inhomogeneous Universe. One aspect of this issue concerns the validity of the continuous approximation: does a system of point masses expand the same way as a fluid does? This article shows that it is not exactly the case in Newtonian gravity, although the associated corrections vanish in an infinite Universe. It turns out that Gauss's law is a key ingredient for such corrections to vanish. Backreaction, therefore, generically arises in alternative theories of gravitation, which threatens the trustworthiness of their cosmological tests. This phenomenon is illustrated with a toy model of massive gravity. 
\end{abstract}

\date{6 June 2017}
\maketitle

\section{Introduction}

A long-standing fundamental question in cosmology is whether a homogeneous and isotropic model accurately predicts the expansion dynamics of our late-time inhomogeneous Universe. The idea that the formation of structures in the cosmos could produce a feedback on cosmic expansion has been formalized by Buchert's work in the late 1990s~\cite{2000GReGr..32..105B}, and is known as the \emph{backreaction problem}~\cite{2011arXiv1109.2314C,2011CQGra..28p4007B,2012ARNPS..62...57B,2013arXiv1311.3787W}. Backreaction can be considered a twofold issue. The first aspect is the one originally discussed by Buchert: due to the nonlinearities of relativistic gravitation, spatial coarse graining and time evolution do not commute; hence, the coarse-grained evolution of an inhomogeneous universe should differ from the evolution of a coarse-grained universe~\cite{1971grc..conf..104E}. This phenomenon does not occur in Newtonian gravity~\cite{1997A&A...320....1B}. The second aspect concerns the validity of the description of matter as a continuous medium---does a fluid universe expand the same way as a universe made of point masses?

While the first aspect is still actively debated, both on
pure-theory~\cite{2013JCAP...12..051L,2015CQGra..32u5013K,2014CQGra..31w4003G, 2015CQGra..32u5021B, 2015arXiv150606452G, 2016CQGra..33l5027G} and numerical bases~\cite{2016PhRvL.116y1301G,2016PhRvL.116y1302B,2016arXiv160708797R,2017arXiv170206643C}, it seems that a consensus has been reached on the second aspect: the fluid approximation seems to be fine. A considerable amount of model-based analyses indeed point towards this direction. Among them can be cited (i)~the Einstein-Straus Swiss-cheese model~\cite{1945RvMP...17..120E,1946RvMP...18..148E,2014JCAP...06..054F}, where spherical regions of a Friedmann-Lema\^itre-Robertson-Walker (FLRW) model can be replaced by point masses without affecting the global expansion law; (ii)~discrete Newtonian cosmologies~\cite{2013arXiv1308.1852G,2009JSMTE..04..019J}; (iii)~finite lattices in GR tessellating~$\mathbb{S}^3$, either based on approximate solutions~\cite{1957RvMP...29..432L,1959RvMP...31..839L,2009PhRvD..80j3503C,2011PhRvD..84j9902C,2015PhRvD..92f3529L,Clifton:2016mxx}, exact solutions~\cite{Clifton:2012qh,Clifton:2013jpa,2014CQGra..31h5002K,Clifton:2014lha,Clifton:2014mza}, or numerical relativity~\cite{Bentivegna:2012ei}; (iv)~infinite cubic lattice universes, again either based on approximate solutions~\cite{2012CQGra..29o5001B}, numerical relativity~\cite{Bentivegna:2013xna,Bentivegna:2013jta,Yoo:2012jz,Yoo:2013yea,Yoo:2014boa}, or most recently on a post-Newtonian expansion~\cite{2015PhRvD..91j3532S,2016PhRvD..94b3505S}.

The above models were naturally worked out in the context of general relativity (GR), which undoubtably represents our best theory of gravitation so far. However, the well-established fact that cosmic expansion is accelerating~\cite{2014A&A...568A..22B,2015arXiv150201589P} stimulated a significant research effort on alternative theories of gravitation which could challenge the cosmological constant as the origin of this acceleration: scalar-tensor theories, including $f(R)$ models~\cite{2010LRR....13....3D}; extended teleparallel models~\cite{2015arXiv151107586C}; massive~\cite{2014LRR....17....7D} and bi-metric theories~\cite{2015PhLB..748...37A}; Lorentz-violating approaches like Ho\v{r}ava-Lifshitz and Einstein-\ae ther theories~\cite{2008arXiv0801.1547J}, etc.---see Ref.~\cite{2012PhR...513....1C} for a comprehensive review. Because cosmology is the main motivation for such alternatives, their confrontation with cosmological data represents an essential test which---this is the key point---is always performed using the homogeneous and isotropic FLRW model in order to deal with cosmic expansion~\cite{2016arXiv160803746D}.

It must be mentioned however that the backreaction issue, or more generally the connection between the small-scale and large-scale behaviors of modified theories of gravity has recently gained more attention. For example it has been shown in Ref.~\cite{Clifton:2015ira}, with a bottom-up approach, that $f(R)$ models which are healthy on small scales tend to have wrong cosmologies and vice-versa. A similar analysis applied to general scalar-tensor and vector-tensor theories has been proposed in Ref.~\cite{0264-9381-34-6-065003}. In Refs.~\cite{2014JCAP...09..017P,2016JCAP...08..038P}, the Green \& Wald approach to backreaction~\cite{2011PhRvD..83h4020G} has been applied to several non-GR theories, leading to a qualitatively different result compared to the GR case.

The present article focuses on the validity of the continuous-medium approximation specifically. It is indeed hard to identify the actual reasons of its success in GR, because because the analyses of discrete cosmological models reported above are all-model based. Extrapolating the accuracy of this approximation to non-GR theories of gravitation is therefore a poorly controlled operation. This leads us to the two central questions addressed in the present article: (i)~Why does a fluid model predicts the right expansion law for a clumpy universe? (ii)~Does this hold in nonstandard theories of gravity?

Hereafter, Sec.~\ref{sec:homogeneous_Newtonian_cosmo} is a quick reminder on how to derive the Friedmann equations in Newtonian cosmology from a purely energetic approach. This draws a path along which I compare, in Sec.~\ref{sec:discrete_Newtonian_cosmo}, the properties of a discrete and a continuous model. This comparison shows that Gauss's law is a key feature of Newtonian gravity for those two models to match. I then repeat in Sec.~\ref{sec:discrete_Yukawa_cosmo} the same analysis with an alternative toy theory of gravitation, namely Yukawa gravity, and discuss the trustworthiness of cosmological tests of modified theories of gravity.

\section{Homogeneous Newtonian cosmology}
\label{sec:homogeneous_Newtonian_cosmo}

Throughout this article, I will adopt Newtonian theory as the standard theory of gravity. It is indeed generally assumed, though debatable, that Newtonian gravity is a good approximation of GR in a cosmological context, provided light is not involved\footnote{In Newton's theory, the gravitational charge is mass rather than energy. This has two direct consequences. First, since light is massless, it cannot fall, which implies no gravitational lensing, in particular no gravitational focusing, hence, wrong predictions for the angular or luminosity distance-redshift relations even in homogeneous cosmology. Second, light cannot be a source of Newtonian gravitation, which makes this theory unable to correctly describe, e.g., the cosmic expansion dynamics during the radiation-dominated era.}. This section is a reminder about Newtonian cosmology, where the Friedmann equation governing the expansion of the Universe can be derived from energetic considerations.

\subsection{Standard energetic approach}

Consider a universe, infinite or not, homogeneously filled with a density~$\rho$ of matter. The symmetries of the system imply that it can expand or contract homothetically, so that $\rho$ is a function of time. Pick an arbitrary origin within this system. For the distribution of matter to remain homogeneous and its flow to be isotropic, the matter velocity field must obey Hubble's law:
\begin{equation}\label{eq:Hubble_law}
\vect{v}(t,\vect{x}) = H(t)\vect{x}.
\end{equation}
If one follows the motion of an arbitrary particle within the flow, its position as a function of time (the displacement field) is then given by
\begin{equation}\label{eq:displacement_field}
\vect{x}(t_2) = \frac{a(t_2)}{a(t_1)}\,\vect{x}(t_1),
\end{equation}
for any two $t_1, t_2$, and where the scale factor $a(t)$ is a function of time such that $\dot{a}/a=H$. 

Consider a spherical region around the origin with constant mass~$M$. It is a closed system, in the sense that we follow the motion of each particle contained in this region; hence, its radius~$r(t)$ evolves according to Eq.~\eqref{eq:displacement_field}. Besides, mass conservation implies $4\pi \rho(t) r^3(t)/3=M$.

As in any closed system, the total (kinetic plus gravitational) energy~$E$ of the ball is conserved with time. It is straightforward to calculate its expression as a function of the radius:
\begin{align}
E &= E\e{kin} + E\e{grav} \\
&=\int \dd^3\vect{x}\;\frac{\rho v^2}{2} - \int \dd^3\vect{x}\,\dd^3\vect{x}'\;\frac{G \rho^2}{\abs{\vect{x}-\vect{x}'}}\\
&= \frac{3}{10} M H^2 r^2 - \frac{3}{5} \frac{G M^2}{r},
\end{align}
whence
\begin{equation}
H^2 = \frac{8\pi G \rho}{3} + \frac{10}{3} \frac{E}{M} \frac{1}{r^2} .
\end{equation}
By taking the present time $t_0$ as a reference, in the sense that $a(t_0)=1$, we can turn the latter equation into a more familiar form: define $K=-10E/[3M r^2_0]$ and get
\begin{equation}
H^2 = \frac{8\pi G\rho}{3} - \frac{K}{a^2},
\label{eq:Friedmann_1}
\end{equation}
which is the first Friedmann equation. Note that for this equation to be independent from the region we started with, i.e. for $K$ to be the same whatever the reference radius~$r_0$, energy cannot be homogeneously distributed in the Universe, more precisely we need~$E\propto r^5_0$. Paradoxically this does not contradict the homogeneity requirement, on the contrary it does ensure it\footnote{When this assumption is broken, we fall into the more general class of models which are spherically symmetric but inhomogeneous, known in GR as the Lema\^itre-Tolman-Bondi (LTB) models---see e.g. Ref.~\cite{2010MNRAS.405.2231F}. The function $-10 E(r_0)/[3M(r_0)r_0^2]$ then corresponds to the spatial curvature function usually denoted $k(r)$ of the LTB solution.}. The second Friedmann equation,
\begin{equation}
\frac{\ddot{a}}{a} = -\frac{4\pi G\rho}{3},
\end{equation}
is then directly obtained by taking the time derivative of Eq.~\eqref{eq:Friedmann_1} and using mass conservation.

Note, finally, that we could have accounted for a cosmological constant by simply adding a term of the form $-M\Lambda r^2/10$ to the gravitational potential energy.\footnote{In a Newtonian context, the cosmological constant cannot be included if gravitation is interpreted as an interaction between massive bodies; it requires a field formulation. The cosmological constant is then an additional, homogeneous, source of the Poisson equation: $\Delta\Phi=4\pi G \rho - \Lambda$.}

\subsection{Discussion}

In the above derivation, the dynamics of cosmic expansion is an exchange between kinetic energy and gravitational potential energy. The deceleration of expansion is the price that must be paid by kinetic energy to gravity which opposes a resistance to it. Such a reasoning equally applies to any self-gravitating system. However, one then has to be careful about the physical meaning of the $E\e{kin}$ and $E\e{grav}$. Indeed, because we are interested in the expansion dynamics only, they must represent respectively the kinetic energy \emph{of expansion} and the gravitational energy that really resists \emph{to this expansion}.

This last point is particularly important in the presence of gravitationally collapsed structures, e.g. galaxies. In such systems, the internal thermal or rotational motion does not contribute to the kinetic energy of expansion, and the gravitational interaction between stars within the galaxy does not resist to the expansion. In other words, the internal degrees of freedom of gravitationally collapsed structures are decoupled from the expansion dynamics. However, the proper distinction between scales which do not participate to the expansion dynamics, and those which do, is highly nontrivial in practice. In the next section I will consider an idealized situation---a discrete cosmological model---in which this distinction is clear, and whose consequences can be quantified. Note that similar issues have been addressed in Ref.~\cite{2012JCAP...07..051B} following a very different approach, based on renormalization techniques.

\section{Discrete Newtonian cosmology}
\label{sec:discrete_Newtonian_cosmo}

Let us compare the energetic properties of a discrete and a continuous gravitational system. We will see that the discreteness of the matter distribution leads to a rescaling of both kinetic and gravitational energy, effectively producing a small change of the spatial curvature term in the expansion dynamics.

\subsection{Description of the models}
\label{subsec:models}

\begin{figure}[h!]
\centering
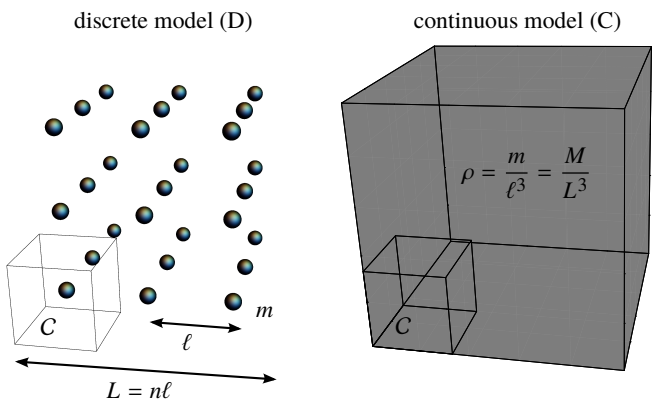
\caption{Two models are compared: a discrete model~\disc\xspace made of~$n^3$ point masses in a cubic lattice, and the corresponding continuous model~\cont\xspace with constant density.}
\label{fig:discrete_model}
\end{figure}

Consider a finite system of $N=n^3$ identical masses~$m$, each one being located at the center of a cell of a cubic lattice of size $L$, as illustrated in Fig.~\ref{fig:discrete_model}. We call $M=Nm$ the total mass and $\ell=L/n$ the distance between two neighboring masses, that is the size of an elementary cell~\cell. This system will be refered to as the discrete cosmological model~\disc\xspace in the following. We also consider the corresponding continuous model~\cont, where the same cells are homogeneously filled with a density $\rho=m/\ell^3$.

The finiteness of those models allows us to easily formulate energetic rationales. They are expected then to behave more similarly to the finite lattice universes studied in Refs.~\cite{1957RvMP...29..432L,1959RvMP...31..839L,2009PhRvD..80j3503C,2011PhRvD..84j9902C,2015PhRvD..92f3529L,Clifton:2016mxx,Clifton:2012qh,Clifton:2013jpa,2014CQGra..31h5002K,Clifton:2014lha,Clifton:2014mza,Bentivegna:2012ei} than to the infinite periodic models of Refs.~\cite{2012CQGra..29o5001B,Bentivegna:2013xna,Bentivegna:2013jta,Yoo:2012jz,Yoo:2013yea,Yoo:2014boa,2015PhRvD..91j3532S,2016PhRvD..94b3505S}. It is thus the opportunity to determine if the backreaction effect observed in finite lattices~\cite{Clifton:2012qh,Bentivegna:2012ei,2014CQGra..31h5002K} is genuinely relativistic, or if it can be captured by Newtonian physics.

\newcommand{\rhoinfigure}{\rho=\dfrac{m}{\ell^3}=\dfrac{M}{L^3}}

\subsection{Effect of discretization on the gravitational potential energy}
\label{subsec:potential_energy}

We start by comparing the gravitational energies of the discrete and continuous models, $E\e{grav}\h{(D)}$ and $E\e{grav}\h{(C)}$. For that purpose, it is convenient to regard model~\cont\xspace as a set of $N$ homogeneous cubes distributed on the same lattice as model~\disc. We can then split the integral of~$E\e{grav}\h{(C)}$ into two contributions: on the one hand, interactions between elements of volume which belong to the same cell~\cell, and, on the other hand, interactions between elements belonging to different cells \cell, \cell':
\begin{multline}
E\e{grav}\h{(C)}
= \overbrace{-\sum_{\mathcal{C}} \int_{\vect{x},\vect{x}'\in\mathcal{C}} \dd^3\vect{x}\,\dd^3\vect{x}' \; \frac{G\rho^2}{\abs{\vect{x}-\vect{x}'}}}^{N E\e{grav,self}} \\
\underbrace{-\sum_{\mathcal{C}\not=\mathcal{C}'} \int_{\vect{x}\in\mathcal{C}, \vect{x}'\in\mathcal{C}'} \dd^3\vect{x}\,\dd^3\vect{x}' \; \frac{G\rho^2}{\abs{\vect{x}-\vect{x}'}}}_{E\e{grav,int}}.
\end{multline}
The first term represents the sum of the gravitational self-energy of each cells, while the second term is the interaction energy between different cells.

Because~$E\e{grav,self}$ and $E\e{grav}\h{(C)}$ represent the gravitational energy of the same system modulo rescaling, we can expect them to be simply related. This can be proved without any calculation but invoking the Buckingham $\Pi$-theorem~\cite{1914PhRv....4..345B}. For the system considered here---a homogeneous cube with size $\ell$ and mass $m$---the dimensionless quantity~$E\e{grav,self}/(G m^2/\ell)$ can only be a function of the other independent dimensionless quantities that one can construct from the parameters of the system and the relevant physical constants. It turns out that there are no such other quantities, which implies that $E\e{grav,self}\propto G m^2/\ell$, whence
\begin{equation}\label{eq:self_grav_Newtonian}
\frac{E\e{grav}\h{(C)}}{E\e{grav,self}}
= \frac{M^2 \ell}{m^2 L} = N^{5/3} .
\end{equation}

Regarding the interaction energy, I will make the approximation that the interaction energy between two different cells $\mathcal{C}$, $\mathcal{C}'$ is equal to the interaction energy of two point masses separated by the same distance, so that
\begin{equation}\label{eq:inter_grav_Newtonian}
E\e{grav,int} \approx E\e{grav}\h{(D)} \define \frac{1}{2} \sum_{i,j\in\text{lattice}} \frac{-G m^2}{\abs{\vect{x}_i-\vect{x}_j}}.
\end{equation}
This is motivated by the so-called \emph{shell theorem} for spherically symmetric distributions, which can be regarded as a consequence of Gauss's law. Although strict equality is only achieved for spherically symmetric distributions, this approximation is shown to be accurate at $5\%$ level — see Appendix~\ref{app:accuracy_cubes_points} for details.

From Eqs.~\eqref{eq:self_grav_Newtonian} and \eqref{eq:inter_grav_Newtonian}, we then conclude that
\begin{equation}\label{eq:ratio_disc_cont}
E\e{grav}\h{(D)} \approx \pa{ 1 - N^{-2/3} } E\e{grav}\h{(C)}.
\end{equation}
In other terms, the gravitational interaction resisting to cosmic expansion is slightly \emph{weaker} in the discrete model than in the continuous model. This must be understood as follows. In the continuous model, every interaction between each element of fluid is involved in this resistance, including interactions between elements that, in the discrete model, would belong to the same body. In the discrete model, there are effectively less gravitational interactions, because internal gravity does not participate. As far as gravitational energy is concerned, the discreteness of the model is equivalent to a renormalization of the gravitational constant as
\begin{equation}
\frac{G\e{eff}}{G} \approx 1 - N^{-2/3} .
\end{equation}
Of course this correction could equivalently be encoded into an effective density~$\rho\e{eff}$ or an effective total mass~$M\e{eff}$.

\begin{figure}[h!]
\centering
\includegraphics[width=\columnwidth]{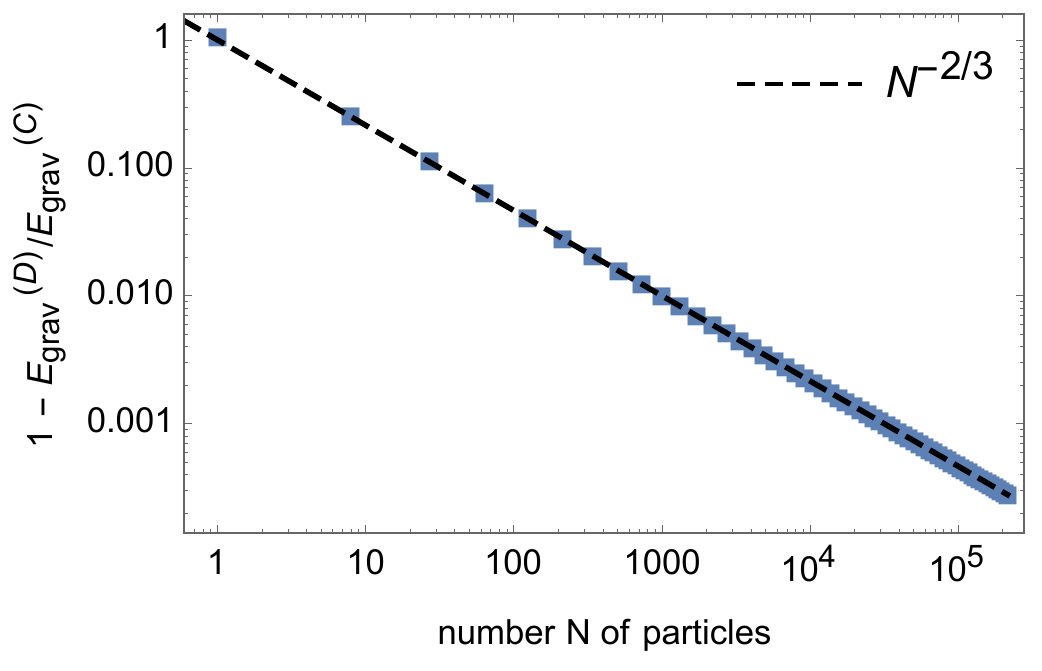}
\caption{Ratio between the gravitational potential energy of the discrete model, $E\e{grav}\h{(D)}$ and of the continuous model, $E\e{grav}\h{(C)}$, as a function of the number~$N$ of particles in the discrete model. Here the total size~$L$ of the lattice is fixed, so that $\ell=L/N^{1/3}$ decreases with $N$. Squares indicate numerical calculations, obtained by computing directly the finite sum of Eq.~\eqref{eq:inter_grav_Newtonian}. They accurately follow the behavior obtained in Eq.~\eqref{eq:ratio_disc_cont}, indicated by the dashed line.}
\label{fig:E_grav_discrete_Newtonian}
\end{figure}

This result has been tested against numerical calculations, as illustrated in Fig.~\ref{fig:E_grav_discrete_Newtonian}. We see that the behavior in $N^{-2/3}$ is accurately reproduced, even for small values of $N$, which confirms the efficiency of the approximation made in Eq.~\eqref{eq:inter_grav_Newtonian}. This power-law correction strongly reminds us the result obtained by Bentivegna and Korzy\'{n}ski~\cite{Bentivegna:2012ei,2014CQGra..31h5002K}, except that the effect seems to go the other way round---their discrete model is effectively heavier than the continuous model. I'll show in subsection~\ref{subsec:discussion_discrete_Newtonian} why our results actually agree.

\subsection{Effect of discretization on the kinetic energy}
\label{subsec:kinetic_energy}

Let us now turn to the kinetic energy. Because the system (even in the continuous case) is not infinite and because its geometry is not spherically symmetric, Hubble's law~\eqref{eq:Hubble_law} should not apply at all times. However, because we are only interested in the effect of discretization and not about finite-size or symmetry effects, we will assume for simplicity that it does. For model~\cont, the kinetic energy is easily calculated as
\begin{align}
E\e{kin}\h{\cont} &= \int \dd^3\vect{x} \; \frac{\rho v^2}{2} \\
							&= \int_{-L/2}^{L/2} \dd x \, \dd y \, \dd z \; \frac{1}{2}\,\rho H^2 (x^2+y^2+z^2) \\
							&= \frac{1}{8} \, M H^2 L^2.
\end{align}
In the discrete case~\disc, we will assume for convenience that $n$ is an odd integer, so that one of the masses is the center of the lattice, with respect to which the other masses are located at positions~$\vect{p}=(p_x,p_y,p_z)$, where each component runs from $-(n-1)/2$ to $(n-1)/2$. The kinetic energy of the system is then
\begin{align}
E\e{kin}\h{\disc} &= \sum_{\vect{p}\in\text{lattice}}  \frac{m v_{\vect{p}}^2}{2} \\
							&= \sum_{p_x,p_y,p_z=-(n-1)/2}^{(n-1)/2} \frac{1}{2} m H^2 \ell^2 (p_x^2+p_y^2+p_z^2) \\
							&= 3 m H^2 \ell^2 \, n^2 \, \frac{n (\frac{n-1}{2})(\frac{n+1}{2})}{6}\\
							&= \frac{1}{8} \, M H^2 L^2 \, \pa{ 1-N^{-2/3} },
\end{align}
from which we conclude that
\begin{equation}
E\e{kin}\h{\disc} = \pa{ 1-N^{-2/3} } E\e{kin}\h{\cont}.
\end{equation}
The discrete model, therefore, has a slightly \emph{lower} kinetic energy of expansion compared to the corresponding continuous model. This can be interpreted as follows. In the continuous model, not only are the cells going away from each other, but they also expand themselves. This second component is absent in the discrete case.

\subsection{Discussion}
\label{subsec:discussion_discrete_Newtonian}

Gathering the results of Subsecs.~\ref{subsec:potential_energy}, \ref{subsec:kinetic_energy} we conclude that, in the discrete model, the relevant energy involved in the expansion dynamics is simply rescaled by a factor $1-N^{-2/3}$ with respect to the continuous case,
\begin{align}
E\e{kin}\h{\disc} + E\e{grav}\h{\disc}
\approx \pa{1-N^{-2/3}} \pa{ E\e{kin}\h{\cont} + E\e{grav}\h{\cont} }.
\end{align}
If this total energy is zero---i.e. in relativistic terms if spatial curvature vanishes---then the dynamics of the discrete model is \emph{identical} to the one of the homogeneous model. If on the contrary the total energy is not exactly zero---if the Universe has a spatial curvature---then the dynamics of model~\disc\xspace differs from the one of~\cont\xspace in that its spatial curvature is effectively weaker:
\begin{equation}
K\h{\disc} \approx \pa{1-N^{-2/3}} K\h{\cont}.
\end{equation}

Physically speaking the scenario is the following. At early times the Universe is very homogeneous, so that $N\rightarrow\infty$ and both curvatures match. As gravitationally bound structures form, $N$ decreases and a fraction $N^{-2/3}$ of both kinetic and gravitational energies leaks from the expansion dynamics to microscopic degrees of freedom, such as the rotational or thermal motions in galaxies and clusters of galaxies, and their gravitational binding energy. The net result of this process is to progressively weaken the corresponding cosmological spatial curvature.

How can this be reconciled with the results of Refs.~\cite{Bentivegna:2012ei,2014CQGra..31h5002K}? Their authors showed that, for a finite lattice of black holes, at the moment of maximum expansion, the expansion dynamics follows the Friedmann equation of a continuum with the effective total mass~$M\e{eff}\approx M(1+N^{-2/3})$. From a Newtonian point of view, at maximum expansion kinetic energy vanishes and total energy is then equal to the effective gravitational energy. We can therefore write
\begin{multline}
\frac{E}{M} \approx \frac{\alpha G M\e{eff}}{L\e{max}} = (1-N^{-2/3}) \times \frac{\alpha G M}{L\e{max}}\\
\Longleftrightarrow (1-N^{-2/3}) \times \frac{E}{M} \approx \frac{\alpha G M}{L\e{max}},
\end{multline}
where $\alpha$ is a geometrical factor. In other words, while Bentivegna and Korzy\'{n}ski work with a fixed spatial curvature (total energy) and interpret the backreaction effect as a enhancement of the mass, I work with a fixed mass and I encode backreaction in a weakened spatial curvature. Both approaches are equivalent at maximum expansion. A similar rationale shows that this behavior in $N^{-2/3}$ also matches the results of Ref.~\cite{Clifton:2012qh}. The remarkable agreement between the results of Refs.~\cite{Bentivegna:2012ei,2014CQGra..31h5002K,Clifton:2012qh} and the calculations of this section shows that, although the latter rely on a fully relativistic treatment of discrete cosmology, the backreaction effect that they observe is, in fact, \emph{Newtonian}.

For reasonable orders of magnitudes of $N$, the associated correction turns out to be very small. For instance, if we suppose that the particles of the discrete model represent clusters of galaxies with mass~$m\sim 10^{15}M_\odot$, then there are approximately~$N\sim 10^8$ such objects in the observable Universe, so that the associated correction of curvature would be on the order of $10^{-6}$. Note finally that such a correction obviously vanishes in an infinite universe ($N\rightarrow \infty$).

The analysis presented in the present section justifies the use of the continuous approximation in Newtonian cosmology. It also emphasizes that its validity is not as trivial as one could expect. A key element in this demonstration is Eq.~\eqref{eq:inter_grav_Newtonian}, i.e. the fact that any isolated distribution of mass gravitates similarly to the same mass concentrated at its center, which I loosely refer to as Gauss's law. This feature is however very specific to fields whose Green function goes like $1/r^2$, hence \emph{it generically does not hold for modified theories of gravitation}, as will be illustrated in the next section.

\section{Discrete cosmology with a modified theory of gravity}
\label{sec:discrete_Yukawa_cosmo}

Several theories of gravitation have been proposed in order to address the astrophysical and cosmological problems of dark matter and dark energy~\cite{2009PhRvD..80j3503C}. Most of those theories are generically expected to violate Gauss's law, so that the conclusions obtained in the previous section cannot apply. Hints towards this statement are, e.g., Ref.~\cite{2008PhRvD..77f4016D} where it is shown that the Dvali-Gabadadze-Porrati gravity~\cite{2000PhLB..485..208D} violates Birkhoff's theorem; or the impossibility of designing Einstein-Straus models with $f(R)$ theories~\cite{Clifton:2012ry}.

In this section, the comparison between a discrete and a continuous model is performed again, but using Yukawa gravity instead of Newtonian gravity. Albeit simplistic compared to the current very elaborated theories of massive gravity~\cite{2014LRR....17....7D},  the Yukawa behavior is known to generically emerge in the weak-field regime of $f(R)$ theories~\cite{2009PhRvD..80j3503C}. Most importantly, this model has the advantage of being easily compared with the Newtonian case, which allows us to keep track of the actual reasons why a backreaction effect could emerge.

\subsection{Yukawa gravity}

The Yukawa theory is obtained by simply adding a mass term to the Poisson equation for the gravitational field~$\Phi$. The resulting field equation takes the form
\begin{equation}\label{eq:Yukawa}
\Delta \Phi - \lambda^{-2} \Phi = 4\pi G \rho,
\end{equation}
where $\rho$ is the local mass density, and $\lambda\define h/(m_\Phi c)$ is the Compton wavelength associated with the mass~$m_\Phi$ of the Yukawa field, $h$ being the Planck constant and $c$ the speed of light.

It is well know that the graviton-mass term in the above equation implies that the field generated by a point mass~$m$ decays exponentially as one goes away from the source:
\begin{equation}
\Phi_\bullet(r) = -\frac{G m}{r} \, \ex{-r/\lambda},
\end{equation}
The exponential factor is sometimes called screening term, by analogy with the behavior of the electrostatic field around individual charges in a plasma or an electrolyte, which are screened by the alternation of successive opposite-charge layers~\cite{DebyeHuckel}.

Integrating Eq.~\eqref{eq:Yukawa} over a spatial domain~$\mathcal{D}$ directly shows that the mass term leads to a violation of Gauss's law,
\begin{equation}
\int_{\partial\mathcal{D}} \vect{g}\cdot\vect{n}\,\dd S = -4\pi G M_\mathcal{D} - \lambda^{-2} \int_{\mathcal{D}} \Phi \; \dd V 
\end{equation}
with $\vect{g}\define -\vect{\nabla}\Phi$, and where $\vect{n}$ is the outwards unit vector normal to the boundary~$\partial\mathcal{D}$ of $\mathcal{D}$. This violation is easily observed if we consider the potential~$\Phi_\ocircle$ created by a sphere of radius $R$ homogeneously filled with density~$\rho$,
%
\begin{equation}
\Phi_\ocircle(r) =
\begin{cases}
4\pi G \lambda^2\rho \pac{ \frac{R+\lambda}{r}\sinh\pa{\frac{r}{\lambda}}\ex{-R/\lambda} - 1}& \text{if }r \leq R\\[3mm]
-\frac{G m}{r} \, \Gamma\pa{\frac{R}{\lambda}} \ex{-r/\lambda} & \text{if }r \geq R
\end{cases}
\label{eq:Yukawa_potential_homogeneous_sphere}
\end{equation}
%
with
\begin{equation}
\Gamma(x) \define \frac{3}{x^3} \pa{ x\cosh x -\sinh x }.
\end{equation}
Given the behaviour of $\Gamma$ at small $x$,
\begin{equation}
\Gamma(x\ll 1) = 1 + \frac{x^2}{10} + \mathcal{O}(x^4),
\end{equation}
one can check that the Newtonian expression is recovered for $\lambda\rightarrow\infty$; in other words there is no vDVZ discontinuity~\cite{1970NuPhB..22..397V,1970JETPL..12..312Z} in Yukawa gravity. In the second line of Eq.~\eqref{eq:Yukawa_potential_homogeneous_sphere}, we see that contrary to Newtonian gravity the gravitational field created outside a homogeneous massive sphere is \emph{not} equivalent to the same mass~$m$ concentrated at its center, but rather to the same mass corrected by a factor $\Gamma(R/\lambda)$.


\subsection{Gravitational interaction of two balls}
\label{subsec:interaction_balls_Yukawa}

\begin{figure}[h!]
\centering
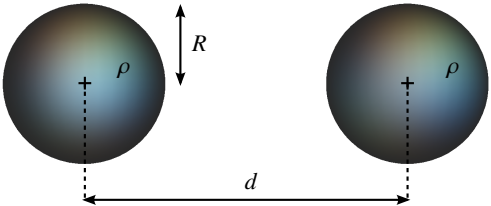
\caption{Two homogeneous balls in gravitational interaction.}
\label{fig:balls}
\end{figure}

Consider two homogeneous balls with density~$\rho$, radius $R$, and whose centers are separated by a distance~$d$, as depicted in Fig.~\ref{fig:balls}. Let us calculate their gravitational interaction energy~$E\e{grav,int}^{\ocircle-\ocircle}$, defined by as the potential energy of one of them, say $\oone$, in the field~$\Phi_{\otwo}$ created by the second one,
\begin{equation}
E\e{grav,int}^{\ocircle-\ocircle}
= \int_{\oone} \dd^3 \vect{x} \; \rho \Phi_{\otwo}(\vect{x}).
\end{equation}

If the balls are disjoint ($d \geq 2 R$), then $\Phi_{\otwo}$ is given by the second line of Eq.~\eqref{eq:Yukawa_potential_homogeneous_sphere}. The integral over the second ball can then be performed analytically and yields
\begin{align}
E\e{grav,int}^{\ocircle-\ocircle}
= -\frac{G m^2}{d} \, \Gamma^2\pa{\frac{R}{\lambda}} \ex{-d/\lambda}.
\end{align}
In other words, the ratio between the interaction energy of two balls and the interaction energy of two point masses of the same mass and separated by the same distance is
\begin{equation}
\frac{E\e{grav,int}^{\ocircle-\ocircle}}{E\e{grav,int}^{\bullet-\bullet}} = \Gamma^2\pa{\frac{R}{\lambda}} .
\end{equation}
It is remarkable that this ratio does not depend on the distance~$d$.

\subsection{Gravitational self-energy of a ball}

In order to repeat a rationale calculation as in the Newtonian case, we need to compute the gravitational self-energy of a homogeneous distribution of mass. In the case of a ball with density $\rho$, this quantity is defined as
\begin{equation}
E\e{grav,self}^\ocircle \define \int_\ocircle \dd^3\vect{x} \; \rho \Phi_\ocircle(\vect{x}),
\end{equation}
where we have to use the first line of Eq.~\eqref{eq:Yukawa_potential_homogeneous_sphere}. Unlike the Newtonian case, since there is a new length scale in the problem, $\lambda$, the $\Pi$-theorem does not directly give the scaling law for $E\e{grav,self}$. We can nevertheless perform the integration analytically and get
\begin{equation}
E\e{grav,self}^\ocircle = -\frac{3}{5} \frac{G m^2}{R} \, \Delta\pa{\frac{R}{\lambda}}.
\end{equation}
The function~$\Delta$, whose expression is
\begin{align}
\Delta(x) &\define \frac{5}{2 x^2} \pac{ 1 - (1+x)\ex{-x} \Gamma(x) } \\
				&= \frac{15}{4 x^5} \pac{ 1 - x^2 + \frac{2 x^3}{3} - (1+x)^2 \ex{-2x} },
\end{align}
quantifies the difference between the Yukawa and Newtonian cases. One can check that the latter is recovered in the limit~$\lambda\rightarrow\infty$, because
\begin{equation}
\Delta(x\ll 1) = 1 - \frac{5 x}{6} + \mathcal{O}(x^2).
\end{equation}

\subsection{Gravitational energy of the discrete universe}

We now turn to the comparison between the discrete~\disc\xspace and continuous~\cont\xspace models presented in Subsec.~\ref{subsec:models}, in the context of Yukawa gravity. We proceed as in the Newtonian case, and start by splitting the total gravitational energy of model~\cont\xspace into the self-energy of individual cells~$\cell$ and the interaction energy of different cells,
\begin{equation}
E\e{grav}\h{\cont} = N E\e{grav,self}^\cell + \sum_{\cell\not=\cell'} E\e{grav,int}^{\cell-\cell'}.
\label{eq:grav_energy_cont_Yukawa}
\end{equation}

\begin{figure}[h!]
\centering
\includegraphics[width=\columnwidth]{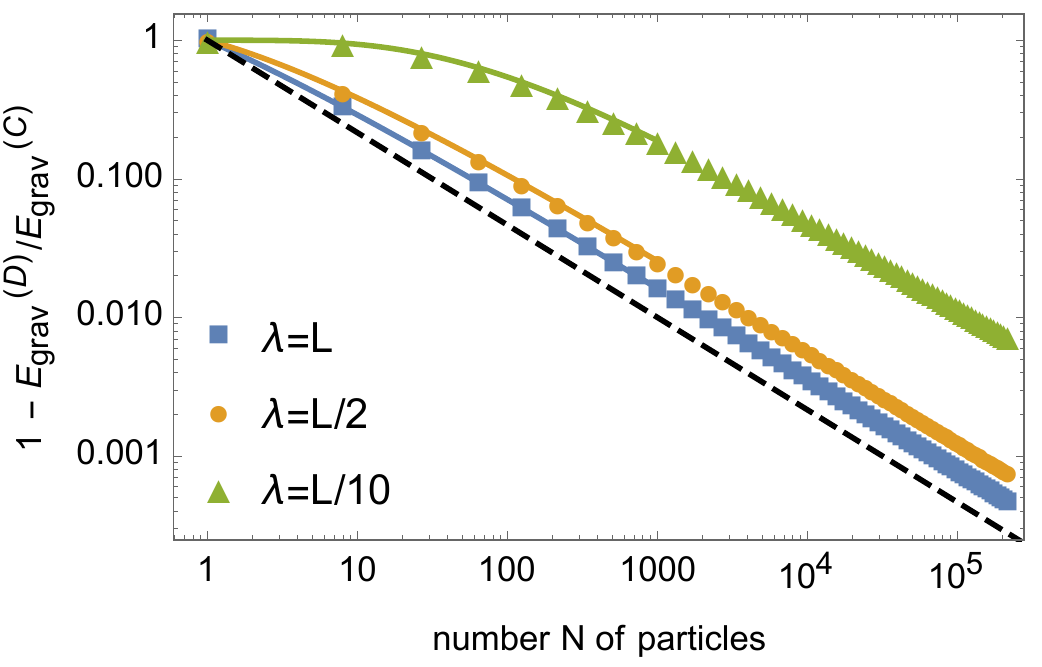}
\caption{Ratio between the gravitational potential energy of the discrete model, $E\e{grav}\h{(D)}$ and of the continuous model, $E\e{grav}\h{(C)}$, as a function of the number~$N$ of particles in the discrete model. Three Yukawa theories are considered, with different masses for the field~$\Phi$, corresponding to Compton lengths $\lambda=L, L/2, L/10$. The total size~$L$ of the lattice is fixed, so that $\ell, R\propto N^{-1/3}$. Squares, disks, and triangles indicate numerical calculations, which follow accurately the behavior predicted by Eq.~\eqref{eq:ratio_disc_cont_Yukawa}, indicated by solid lines. The Newtonian case~$N^{-2/3}$ is indicated by a dashed line for comparison.}
\label{fig:E_grav_discrete_Yukawa}
\end{figure}

We then make the following approximations: although the cells have a cubic geometry, we calculate their self-energy as if they were balls with the same mass~$m$ and volume~$\ell^3$, i.e. with radius
\begin{equation}
R = \pa{\frac{3}{4\pi}}^{1/3} \, \ell,
\end{equation}
in other words,
\begin{equation}
E\e{grav,self}^\cell \approx E\e{grav,self}^\ocircle(m,R).
\end{equation}
In a similar way, we calculate the gravitational energy of the full lattice, i.e. $E\e{grav}\h{\cont}$, as if it were a ball with total mass $M=Nm$ and volume $L^3$, so that
\begin{align}
E\e{grav}\h{\cont} 
&\approx E\e{grav,self}^\ocircle (N m, N^{1/3}R) \\
&\approx N^{5/3} \frac{\Delta(N^{1/3}R/\lambda)}{\Delta(R/\lambda)} E\e{grav,self}^\cell,
\label{eq:self_grav_Yukawa}
\end{align}
which replaces the scaling law~\eqref{eq:self_grav_Newtonian} obtained in the Newtonian case. Finally, we evaluate the interaction of different cells as if there were disjoint balls, which allows us to use the result of Subsec.~\ref{subsec:interaction_balls_Yukawa},
\begin{align}
\sum_{\cell\not=\cell'} E\e{grav,int}^{\cell-\cell'} 
&\approx \sum_{\ocircle\not=\ocircle'} E\e{grav,int}^{\ocircle-\ocircle'} \\
&= \Gamma^2\pa{\frac{R}{\lambda}} \sum_{\bullet\not=\bullet'} E\e{grav,int}^{\bullet-\bullet'} \\
&= \Gamma^2\pa{\frac{R}{\lambda}} E\e{grav}\h{\disc}.
\label{eq:inter_grav_Yukawa}
\end{align}
We conclude from Eqs.~\eqref{eq:grav_energy_cont_Yukawa}, \eqref{eq:self_grav_Yukawa}, and \eqref{eq:inter_grav_Yukawa} that the ratio between the gravitational energies of the two models reads
\begin{equation}\label{eq:ratio_disc_cont_Yukawa}
\frac{E\e{grav}\h{\disc}}{E\e{grav}\h{\cont}}
\approx
\frac{1}{\Gamma^2(R/\lambda)} \pac{ 1- \frac{1}{N^{2/3}} \frac{\Delta(R/\lambda)}{\Delta(N^{1/3}R/\lambda)} }.
\end{equation}
Although the assumptions formulated above can seem crude, they actually provide an excellent approximation of the final result, as illustrated in Fig.~\ref{fig:E_grav_discrete_Yukawa}. Just like in the Newtonian case, Eq.~\eqref{eq:ratio_disc_cont_Yukawa} can be seen as the ratio between the effective gravitational constant in the discrete model and the true gravitational constant.

\subsection{Discussion}

There are two significant distinctions between the Yukawa case analyzed here and the Newtonian case presented in the previous section.

First, the respective changes in gravitational energy and kinetic energy due to discretization are no longer identical. Hence, while the net consequence of the formation of gravitationally bound structures on cosmic expansion was found to be, in the Newtonian case, a rescaling of spatial curvature, this is no longer true in Yukawa gravity. A consequence is that backreaction now occurs also in a Universe with zero spatial curvature.

\begin{figure}[!h]
\centering
\includegraphics[width=\columnwidth]{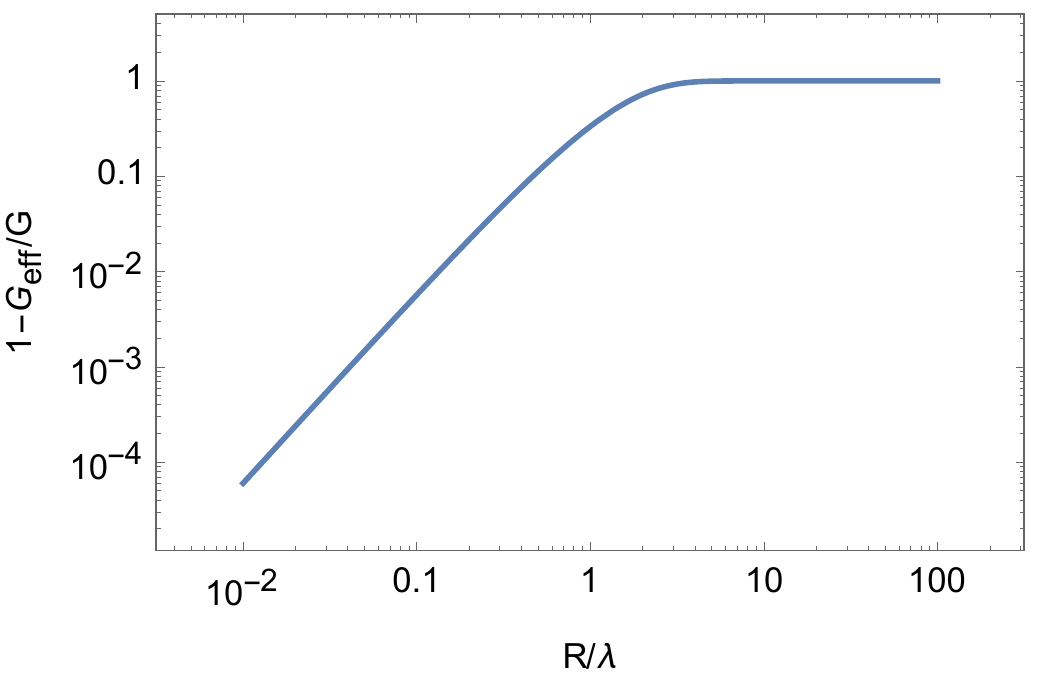}
\caption{Effective gravitational constant~$G\e{eff}$ for the expansion dynamics of an infinite discrete Universe with Yukawa gravity.}
\label{fig:Geff_Yukawa}
\end{figure}

Second and most importantly, with Yukawa gravity \emph{the backreaction effect holds in an infinite Universe}. Indeed, in the limit~$N\rightarrow\infty$ ($R$ finite), the ratio of gravitational energies in the discrete and continuous models, i.e. the ratio between effective and real gravitational constants, reads
\begin{align}\label{eq:Geff_Yukawa_infinite}
\lim{N}{\infty} \frac{G\e{eff}}{G}
&= \frac{1}{\Gamma^2(R/\lambda)} \pac{ 1 - \frac{2}{5}\pa{\frac{R}{\lambda}}^2 \Delta\pa{ \frac{R}{\lambda} } } \\
&\underset{R\ll\lambda}{\approx} 1 - \frac{3}{5}\pa{\frac{R}{\lambda}}^2.
\end{align}
Recall that $R$ is essentially the size of a cell of the discrete model. Physically speaking, it thus represents the typical distance between e.g. two galaxy clusters, or equivalently the size that would have a cluster if its density were the mean density of the Universe, that is~$R\sim 20\U{Mpc}$. If gravitation is assumed to depart from the Newtonian behavior on cosmological scales only, then $\lambda\sim H_0^{-1}$, and we find $(R/\lambda)^2\sim 10^{-6}$, which is a negligible correction. However, if it is modified on scales smaller than $R$, then $G\e{eff}$ goes quickly to zero, as shown in Fig.~\ref{fig:Geff_Yukawa}. Note that this last property is specific to Yukawa theory, because it is short-ranged; albeit a backreaction effect is expected in general, it does not necessarily lead to $G\e{eff}/G\rightarrow 0$ for other alternative theories of gravity.

Let me stress that the correction that we are talking about here is \emph{not} the usual cosmological consequence of modifying our theory of gravity. Such an effect indeed is already taken into account in a homogeneous model. The backreaction effect encapsulated in, e.g., Eq.~\eqref{eq:Geff_Yukawa_infinite} is an \emph{additional} change in the expansion dynamics, which is not accounted for in a model where matter is described as a continuous fluid\footnote{This must be taken as a proof of principles, because in fact a homogeneous Yukawa universe does not gravitate~\cite{Yukawa_cosmo}}.

\section{Conclusion}

The question of whether a strictly homogeneous and isotropic model for the Universe is capable of predicting its expansion dynamics accurately is fundamental in cosmology. One peculiar aspect of this question concerns the validity of the fluid limit, i.e. whether a universe continuously filled with matter behave identically to a universe filled with matter clumps. In this article, I investigated this issue by comparing two models of a finite Universe: in the first one matter is distributed on a lattice; in the second one, it homogeneously fills the box.

In Newtonian gravity, the net effect of the discreteness of the distribution of matter is shown to slightly weaken spatial curvature compared to the homogeneous case, by a factor $1-N^{-2/3}$ where $N$ is the number of particles in the model. This result agrees with earlier results for finite lattice cosmologies in GR~\cite{Bentivegna:2012ei,2014CQGra..31h5002K,Clifton:2012qh}, showing that the corresponding backreaction effect is actually Newtonian. This correction vanishes in an infinite universe, and is very small ($< 10^{-6}$) in a realistic finite universe. It also appeared that such a result crucially relies on the fact that any isolated distribution of mass gravitates similarly to a point mass, which can be regarded as a consequence of Gauss's law.

Gauss's law is very specific to Newtonian gravity and GR.\footnote{Formulating an equivalent of Gauss's law in GR is not trivial, but the possibility of constructing Swiss-cheese models with either Schwarzschild~\cite{2014JCAP...06..054F}, LTB~\cite{2007JCAP...02..013B,2007PhRvD..76l3004M}, or Szekeres~\cite{2009GReGr..41.1737B} holes is a convincing indication that such a law should exist.} It is generically violated in alternative theories, such as massive gravity or $f(R)$ theories. I illustrated this phenomenon with the simple example of Yukawa gravity, characterized by an exponential suppression of gravitational interactions beyond distances controlled by the graviton's Compton length~$\lambda$. The difference between discrete and continuous cosmologies turns out to be qualitatively different from the Newtonian case: first, it does not only lead to a renormalization of spatial curvature, and second the corrections hold in an infinite Universe. The expansion law of a clumpy Universe is expected to significantly differ from the predictions of the homogeneous model if $\lambda\lesssim 20\U{Mpc}$. Albeit still allowed by gravitational wave experiments~\cite{2016PhRvL.116f1102A}, such a value of $\lambda$ would also imply strong deviations from Newtonian gravity on the scale of galaxy clusters, and significant changes in the cosmic structure formation, which is excluded by observations~\cite{PhysRevD.75.044004,2015Univ....1..123D}.

The above results generally question the trustworthiness of the cosmological tests of any modified theory of gravity that violates Gauss's law. Indeed, in such a situation the expansion dynamics predicted by a Friedmann-Lema\^{i}tre-Robertson-Walker model does not match the actual gravitational dynamics of the late-time, structured Universe. The amplitude of this mismatch must be estimated for any analysis of the cosmological data in the scope of modified theories of gravity.

\acknowledgements

I thank Jean-Philippe Uzan for preliminary discussions at early stages of this project, Xiaodong Xu for pointing out the effect of discretization on kinetic energy, and Lorenzo Reverberi for stimulating debates. I am also grateful for comments from Thomas Buchert, Pedro Ferreira, Tim Morris, Cliff Taubes, and especially Timothy Clifton whose great knowledge of the literature made the bibliography much more comprehensive. Finally I would like to thank the anonymous referees for relevant comments which improved the quality of the article, in particular through the addition of the Appendix. The financial assistance of the South African National Research Foundation (NRF) towards this research is hereby acknowledged. Opinions expressed and conclusions arrived at, are those of the author and are not necessarily to be attributed to the NRF.

\appendix
\section{On the accuracy of Eqs.~\eqref{eq:inter_grav_Newtonian} and \eqref{eq:ratio_disc_cont}}
\label{app:accuracy_cubes_points}

In this appendix, we evaluate the accuracy of the approximation stated in Eq.~\eqref{eq:inter_grav_Newtonian}, namely that the gravitational interaction energy of a system of cubic masses is essentially equal to the gravitational energy of a system of point masses,
\begin{equation}
E\e{grav,int} \approx E\e{grav}\h{\disc}.
\end{equation}
We will also analyze the consequences of the small corrections to Eq.~\eqref{eq:inter_grav_Newtonian} on Eq.~\eqref{eq:ratio_disc_cont}, which concerns the ratio between the gravitational energies of the discrete lattice model, $E\h{\disc}\e{grav}$, and its continuous counterpart~$E\h{\cont}\e{grav}$,
\begin{equation}
E\h{\disc}\e{grav} \approx \pa{ 1-N^{-2/3} } E\h{\cont}\e{grav}.
\end{equation}

\begin{figure}[h!]
\centering
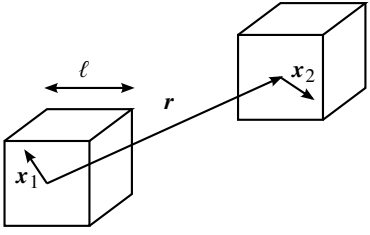
\caption{Two homogeneous cubes separated by $\vect{r}$ in gravitational interaction.}
\label{fig:cubes}
\end{figure}

For that purpose, we start by comparing the gravitational interaction energy of two homogeneous cubes~$E\e{grav,int}^{\cell-\cell}$ with that of two point masses,~$E\e{grav,int}^{\bullet-\bullet}$, with the same mass~$m$ and located at the center of the cubes. We still denote with~$\ell$ the size of the cubes, while~$\vect{r}$ is the relative position between their centers, as illustrated in Fig.~\ref{fig:cubes}. By definition, the gravitational interaction energy of the cubes reads
\begin{equation}\label{eq:interaction_two_cubes}
E\e{grav,int}^{\cell-\cell} = - \frac{G m^2}{\ell^6} \int_{\cell^2} \frac{\dd^3\vect{x}_1 \dd^3\vect{x}_2 }
																												{\abs{\vect{r}-(\vect{x}_2-\vect{x}_1)}},
\end{equation}
where $\vect{x}_{1,2}$ represent positions within each cube with respect to their centers, hence $x^i_{1,2}$ run from $-\ell/2$ to $\ell/2$.

The integrand of Eq.~\eqref{eq:interaction_two_cubes} can be expanded over multipoles as
\begin{equation}
\frac{1}{\abs{\vect{r}-(\vect{x}_2-\vect{x}_1)}}
=
\sum_{k=0}^\infty \frac{\abs{\vect{x}_2-\vect{x}_1}^k}{r^{k+1}} \,
								P_k\pac{ \frac{\vect{r}\cdot(\vect{x}_2-\vect{x}_1)}{r \abs{\vect{x}_2-\vect{x}_1}} },
\end{equation}
with $r\define\abs{\vect{r}}$ and where $P_k$ is the $k$th Legendre polynomial. Performing the integration over each multipole, one obtains that the multipoles~$k=1,2,3$ vanish identically, so that the first term (apart from the monopole) in the expansion is the hexadecapole~$k=4$, as we could except from symmetry considerations. More precisely, we obtain
\begin{equation}\label{eq:hexadecapole}
E\e{grav,int}^{\cell-\cell}
=
- \frac{G m^2}{r} \bigg[ 1 + \alpha_4(\vect{r}) \pa{\frac{\ell}{r}}^4  + \mathcal{O}\pa{\frac{\ell^6}{r^6}} \bigg],
\end{equation}
with
\begin{equation}
\alpha_4(\vect{r}) \define \frac{7}{480} \pa{ 3 - 5 \, \frac{x^4+y^4+z^4}{r^4} },
\end{equation}
where $x,y,z$ denote the components of~$\vect{r}$.

We now have to sum over all couples~$(i,j)$ of cubes. The expansion~\eqref{eq:hexadecapole} will actually serve to show that the difference between cubes and point masses is essentially due to the interaction between closest neighbors. The total interaction energy of the system of cubes reads
\begin{align}
E\e{grav,int} 
&= \frac{1}{2}\sum_{i,j} E^{\cell_i-\cell_j}\e{grav,int} \\
&= E\e{grav}\h{\disc}
 - \frac{1}{2}\frac{Gm^2}{\ell} \sum_{i,j} \Bigg[ \alpha_4(\vect{r}_{ij}) \pa{ \frac{\ell}{r_{ij}} }^5 \nonumber \\
 	& \hspace{4cm} + \mathcal{O}\pa{\frac{\ell^7}{r_{ij}^7}} \Bigg].
\end{align}
An upper bound of the hexadecapole correction can be evaluated the following way. First,
\begin{equation}
\abs{ \sum_{i,j} \alpha_4(\vect{r}_{ij}) \pa{ \frac{\ell}{r_{ij}} }^5}  
\leq \alpha_4\h{max} \sum_{i,j} \pa{ \frac{\ell}{r_{ij}} }^5 ,
\end{equation}
then consider that there are typically~$n(r)\approx\pi (r/\ell)^2$ cubes at a distance~$r$ from any cube of the system, except for cubes close to the boundary of the system, in which case there are even less. We use this property to write the above sum as a sum over $r$, which we then turn into an integral,
\begin{align}
\sum_{i,j} \pa{ \frac{\ell}{r_{ij}} }^5
&= \sum_i \sum_{r_{ij}} n(r_{ij}) \pa{ \frac{\ell}{r_{ij}} }^5 \\
&\approx \sum_i \int_{\ell}^L \pi \pa{ \frac{\ell}{r} }^3 \frac{\dd r}{\ell} \\
&= \frac{N\pi}{2} \pa{ 1 - \frac{\ell^2}{L^2} } \\
&\approx \frac{N\pi}{2},
\end{align}
since in the physically interesting cases~$\ell\ll L$. We conclude that only the small values of $r$ in the sum/integral really contribute. In other words, the difference between~$E\e{grav,int}$ and $E\h{\disc}\e{grav}$, the contribution of all the couples of cubes which are not very close to each other is negligible. The resulting difference,
\begin{equation}
\abs{ E\e{grav,int} - E\e{grav}\h{\disc} } \leq \frac{N\pi \alpha_4\h{max}}{4} \, \frac{G m^2}{\ell},
\end{equation}
indeed corresponds to the contribution of the first layers around each cube, which explains why it scales like $N$. Note however that the above estimation of this contribution is quite crude, because it relies on a multipole expansion which is valid only when $(\ell/r)^7 \ll 1$, it is therefore inaccurate for $r=\ell$ which is the case for closest neighbours. Let us therefore calculate this correction more precisely.

Since we have shown that only the closest interaction must be considered in the difference between $E\e{grav,int}$ and $E\e{grav}\h{\disc}$, we will repeat the calculation of this difference considering only the first layer (26 cubes) around each cube, i.e. assuming that the next layers are equivalent to point masses. Numerically, this correction reads
\begin{align}
\delta E_1
&\define E\e{grav,int}\h{27 cubes} - E\e{grav,int}\h{27 points} \\
&= 8.65 \times 10^{-2} \, \frac{G m^2}{\ell},
\end{align}
from which we deduce
\begin{align}
E\e{grav,int} - E\e{grav}\h{\disc} 
&= \frac{N}{2} \delta E_1 \\
&=\Delta_1 \, N^{-2/3} E\e{grav}\h{\cont},
\end{align}
with $\Delta_1=4.60\times 10^{-2}$. Repeating the calculation of Subsec.~\ref{subsec:potential_energy}, we find
\begin{equation}\label{eq:ratio_discrete_continuous_first_layer}
\frac{E\h{\disc}\e{grav}}{E\h{\cont}\e{grav}} = 1 - (1+\Delta_1)\, N^{-2/3},
\end{equation}

\begin{figure}[t]
\centering
\includegraphics[width=\columnwidth]{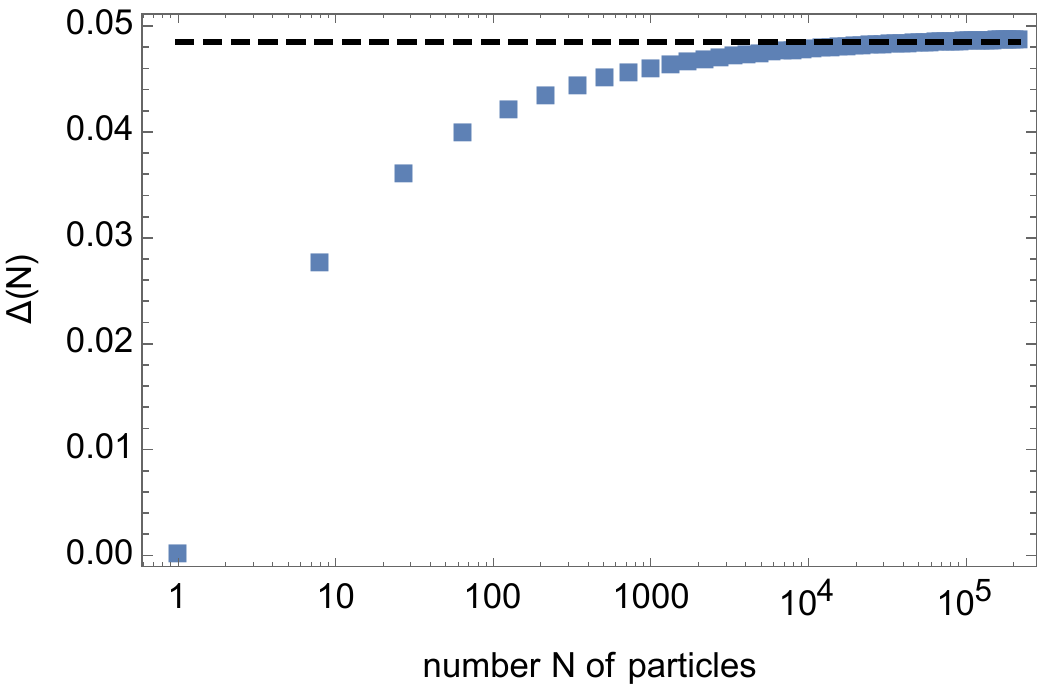}
\caption{Fractional difference~$\Delta$ between $E\h{\disc}\e{grav}/E\h{\cont}\e{grav} - 1$ and the $N^{-2/3}$ law, as defined in Eq.~\eqref{eq:fractional_correction_to_N23_law}, as a function of the number of particles~$N$ in the model. Blue squares indicate numerical calculations, while the dashed line indicates the asymptotic value $\Delta_\infty=4.85\times 10^{-2}$.}
\label{fig:correction_interaction_energy}
\end{figure}

This last result can finally be checked numerically by plotting, in Fig.~\ref{fig:correction_interaction_energy}, the quantity
\begin{equation}\label{eq:fractional_correction_to_N23_law}
\Delta(N) \define N^{2/3} \pa{ \frac{E\h{\disc}\e{grav}}{E\h{\cont}\e{grav}} - 1 + N^{-2/3} },
\end{equation}
that is the fractional difference to the $N^{-2/3}$ law, as a function of $N$. We see that for a large number of masses, $\Delta(N)$ tends to $\Delta_\infty = 4.85 \times 10^{-2}$, which is in good agreement with the estimation~$\Delta_1$ taking only the first layer into account.

Summarizing, the approximation stated in Eq.~\eqref{eq:inter_grav_Newtonian} is accurate at the level of $\sim 5\%$. The corrections  manifest, at large~$N$, as a slight modification of the prefactor of the $N^{-2/3}$ law. Physically speaking, the difference is located at the immediate vicinity of each cell, where the interaction with the closest neighbors is accurate at a $5\%$ level. Although the above calculation has been performed for a cubic lattice, the very nature of this $5\%$ correction implies that it would hold for any kind of lattice, regardless of its geometry.

\bibliography{bibliography_backreaction.bib}

\end{document}